\documentclass[twocolumn]{revtex4}
\usepackage[latin1]{inputenc}
\usepackage[T1]{fontenc}
\usepackage{times}
\usepackage{amsmath}
\usepackage{amsfonts}
\usepackage{amssymb}
\usepackage{algorithm}
\usepackage{algorithmic}
\usepackage{times}
\usepackage{rotating}
\usepackage{multirow}
\usepackage{url}
\usepackage[small,compact]{titlesec}
\usepackage[small,it]{caption}
\begin{document}
\title{Immune System Inspired Strategies for Distributed Systems}
\author{Soumya Banerjee\(^{\S}\)* and Melanie Moses\(^\S\)\\
\(^\S\) Department of Computer Science, University of New Mexico\\}

\begin{abstract}
Many components of the IS are constructed as modular units
which do not need to communicate with each other such that the number of components increases but the size remains constant. However, a sub-modular IS architecture in which lymph node number and size both increase sublinearly with body size is shown to efficiently balance the requirements of communication and migration, consistent with experimental data. We hypothesize that the IS architecture optimizes the tradeoff between local search for pathogens and global response using antibodies. Similar to natural immune systems, physical space and resource are also important constraints on Artificial Immune Systems (AIS), especially distributed systems applications used to connect low-powered sensors using short-range wireless communication. AIS problems like distributed robot control will also require a sub-modular architecture to efficiently balance the tradeoff between local search for a solution and global response or proliferation of the solution between different components.
\end{abstract}
\maketitle
\section{Introduction}
Distributed systems are becoming increasingly important in areas like environmental monitoring, disaster relief, military operations, multi-robot control, mobile ad-hoc networks, intrusion detection systems and malware detection systems (Kleinberg 2007). Such systems typically operate under constraints of physical space and resource (power, bandwidth), and performance scalability is a desirable feature. The natural immune system (NIS) also operates under similar constraints of physical space and resource. The NIS has very rare antigen-specific immune system (IS) cells which it uses to search through the physical space of the whole body to find small amounts of spatially localized antigen. Our empirical data indicates that the time to detect and respond to antigen is still invariant with host body size (Banerjee and Moses 2009, Banerjee and Moses 2010). To facilitate this search, the NIS uses anatomical structures called lymph nodes (LN). A LN is a place in which IS cells, antigen-presenting cells and antigen encounter each in a small local region of tissue. They form a decentralized detection network and are distributed throughout the body. The present work draws inspiration from the architecture of the lymphatic system and discusses architectural strategies for distributed systems.

\begin{figure*}
\centering
\scriptsize
\begin{tabular}{l}
\includegraphics[scale=0.6]{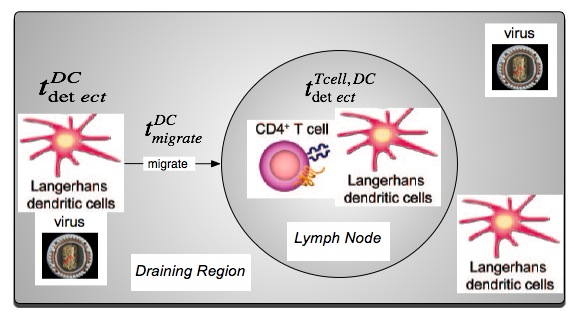}\\
\end{tabular}
\caption{Immune system dynamics within a lymph node and its draining region} 
\label{fig:goslim}
\end{figure*}

\section{Immunological Preliminaries}
The area of tissue that drains into a lymph node (LN) is called its draining region (DR). Dendritic cells (DCs) sample the tissue in DRs for pathogens, and upon encountering them, migrate to the nearest LN T cell area to present antigen to T helper cells. Cognate T helper cells specific to a particular pathogen are very rare (1 in \(10^{6}\) IS cells). Upon recognizing cognate antigen on DCs, T helper cells proliferate and build up a clonal population in a process called clonal expansion. While proliferating, T helper cells also migrate to the LN B cell area to activate B cells. Cognate B cells specific to a particular pathogen are also very rare. They need to recognize cognate antigen on follicular dendritic cells (FDC) and also need stimulation from cognate T helper cells. After recognition, cognate B cells undergo clonal expansion and differentiate into antibody-secreting plasma cells. A diagram detailing a subset of these processes is shown in Fig. 1.

\section{Local Detection and Global Response}
A horse 25,000 times larger than a mouse must generate 25,000 times more absolute quantities of antibody (\(Ab\)) in order to achieve the same concentration of antibody in the blood (where blood volume is proportional to \(M\) (host body mass), Peters et al. 1983). A fixed antibody concentration is required to fight infections like WNV that spread systemically through the blood and hence \(Ab\) is proportional to \(M\) (host body mass).

In all pathogens which evoke the adaptive IS, the rate limiting step is the recognition of antigen on DCs by antigen-specific T helper cells within the LN T cell area (Soderberg et al. 2005). The time taken in this recognition step impacts other downstream processes like activation of B cells since T helper cells activated by DCs must migrate to the B cell area to activate B cells. If organisms of all body sizes activated the same number of B cells, the time for a fixed number of B cells to produce \(Ab\) is proportional to \(log_{2} M\)   (since B cells reproduce exponentially through clonal amplification). For example, since it takes 4 days of exponential growth of activated B cells to produce sufficient anti-WNV neutralizing antibody in mice (Diamond et al. 2003), then the corresponding time for a horse would be more than 2 months. This conflicts with empirical data on horses (Michel et al. 2002). We assume that the IS of larger organisms has to activate a number of antigen-specific B cells (\(B_{crit}\)) proportional to \(M\), in order to build up the critical density of antibodies in a fixed period of time (scale-invariant response: Banerjee and Moses 2009, Banerjee and Moses 2010).

Since only a fixed number of antigen-specific B cells reside inside the LN (1 in \(10^{6}\) IS cells, Soderberg et al. 2005), this implies that as the organism size increases, the infected site LN has to recruit increasing numbers of B cells from other LNs. This necessitates the LN growing larger in larger organisms, since a larger LN will have more high endothelial venules (HEV) to recruit more IS cells. However having a larger LN means that the volume of the DR it services will be very large (since the total amount of lymphoid tissue is proportional to body mass), which will increase the average time taken by antigen-loaded DCs to reach the LN. Hence we see that there is a tradeoff between having a larger LN (reduces global communication time involved in recruiting additional IS cells to ensure a global antibody response) and a smaller LN (reduces local communication time involved in trafficking antigen-loaded DCs to draining LN). In other words, the NIS thinks locally but acts globally.

Analysis of empirical data (Banerjee and Moses 2009, Banerjee and Moses 2010) shows that LN size increases with body mass, indicating that the NIS balances the tradeoff between local and global communication by successively increasing LN size and number of LNs in larger organisms.

\section{Architecture of the NIS}
	In order to explain the empirical data, we explored three possible architectures of lymphatic system organization (diagrammed in Fig. 2).

\begin{figure*}
\centering
\scriptsize
\begin{tabular}{l}
\includegraphics[scale=0.45]{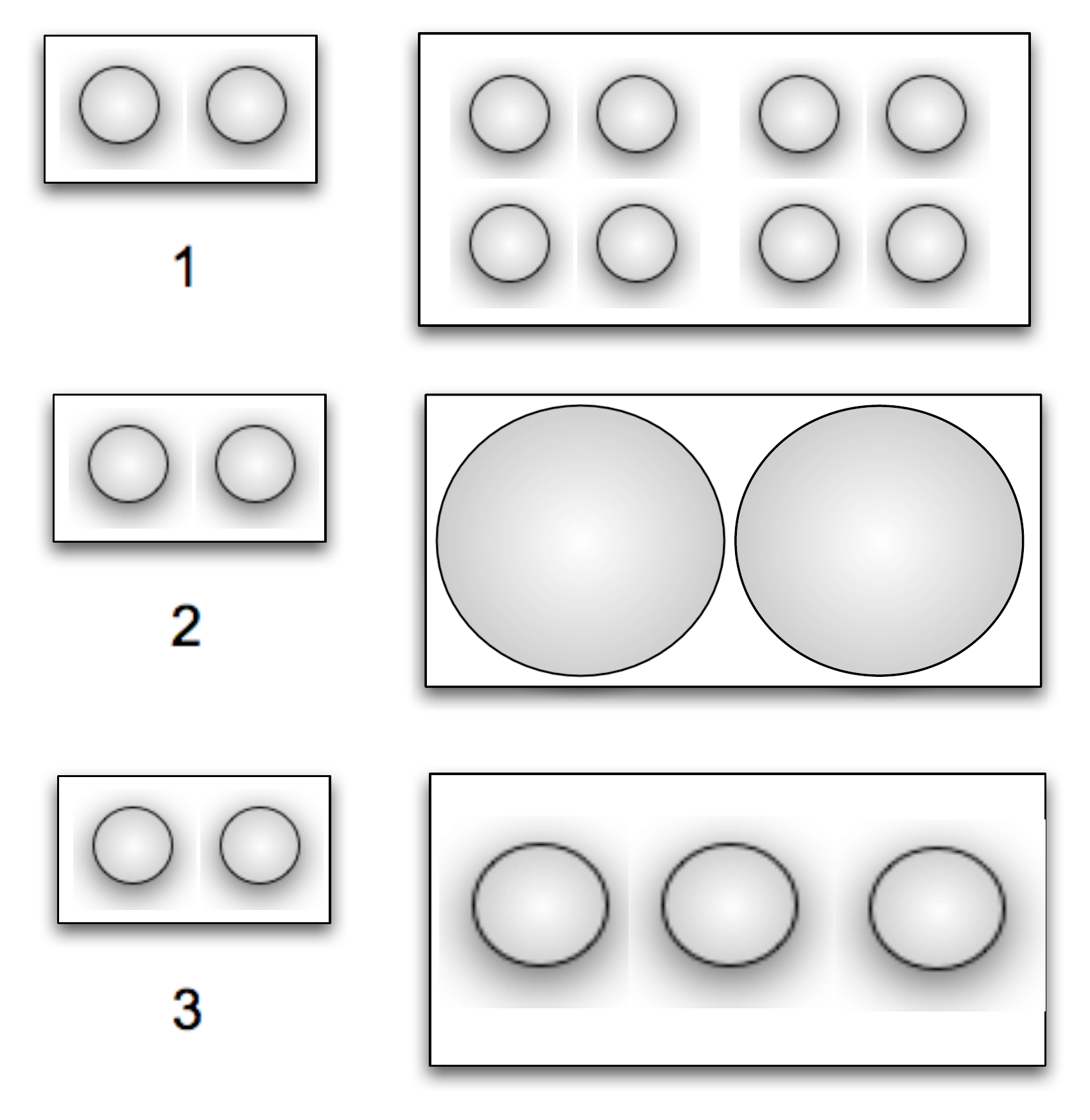}\\
\end{tabular}
\caption{The 3 different hypotheses of scaling of lymph node size and numbers. (1) Completely Modular Detection Network: base organism with 2 LNs and another organism 4 times as big with 4 times the number of lymph nodes each of the same size as the base organism. (2) Non-Modular Detection Network: organism 4 times bigger has the same number of LNs but each is 4 times bigger. (3) Hybrid Sub-Modular Detection Network: organism 4 times bigger has more LNs each of which are also bigger.} 
\label{fig:goslim}
\end{figure*}

1) Model 1 (Completely Modular Detection Network): In a completely modular detection network, the number of LNs increases linearly with system size but the size of each LN remains the same. Each LN has a fixed size DR and hence the average time taken for antigen-loaded DCs to migrate to the draining LN remains the same for organisms of all sizes. However the infected site LN has to incur increasing overheads of recruiting IS cells from an increasing number of LNs (proportional to \(M\)) in order to activate a critical number of B cells (\(B_{crit}\)). Hence the completely modular detection network is optimized for local communication at the expense of global communication.

2) Model 2 (Non-Modular Detection Network): The second model is the other extreme, that LNs are arranged in a detection network with a constant number of LN (all animals have the same number of LNs, however the size of LNs is larger in larger animals). This architecture compensates for the limitation of physically transporting IS cells over larger distances by making LNs bigger (and situating more cognate IS cells in the LN) in larger organisms. However, the size of the DR increases with organism size and hence this architecture incurs increasing local communication costs (time for antigen-loaded DCs to reach draining LN), while there is a fixed global communication cost (since all the necessary IS cells which need to be activated are within the LN).

3) Model 3 (Hybrid Sub-Modular Architecture): In this model LNs increase in both size and in numbers as animal size increases, and so does the size of the DR. This architecture strikes a balance between the two opposing goals of antigen detection (local communication) and antibody production (global communication), and simultaneously minimizes antigen detection time and antibody production time.

	In summary, due to the requirement of activating increasing number of IS cells for antibody production in larger organisms, there are increasing costs to global communication as organisms grow bigger. The semi-modular architecture (Model 3) balances the opposing goals of detecting antigen using local communication and producing antibody using global communication and leads to optimal antigen detection and antibody production time.

\section{Applications to Distributed Systems}
	The natural immune system (NIS) utilizes an architecture which functions within constraints imposed by physical space. Physical space is also an important constraint on artificial immune systems (AIS), especially distributed systems applications (Kleinberg 2007). Such networks are being increasingly used in environmental monitoring, disaster relief, military operations, multi-robot control, mobile ad-hoc networks, intrusion detection systems and malware detection systems (Kleinberg 2007). These networks might operate under constraints similar to an NIS and hence the design of the AIS can be informed by architectural strategies employed by their biological counterpart. 
	
	As a concrete example of an application where space is a constraint and scaling of performance with system size is an important design criterion, we consider an AIS approach to control multiple robots tasked with obstacle avoidance (Shivashankar et al. 2008). The robots communicate with software agent(s) in a server upon encountering an obstacle. The agents transmit rule-sets of actions to robots to help overcome their obstacles, and agents also share information globally amongst themselves by migrating to other computer servers. Some analogies between this AIS and an NIS are: the obstacle problem presented by a robot is an antigen, the rule-set of actions transmitted by an agent are antibodies, the robots are DCs, software agents are B cells, the computer servers themselves are LNs, and the physical area serviced by a single computer server is a DR. The system is diagrammed in Fig. 3 and Fig. 4 (modified from Shivashankar et al. 2008). 
	
\begin{figure*}
\centering
\scriptsize
\begin{tabular}{l}
\includegraphics[scale=0.45]{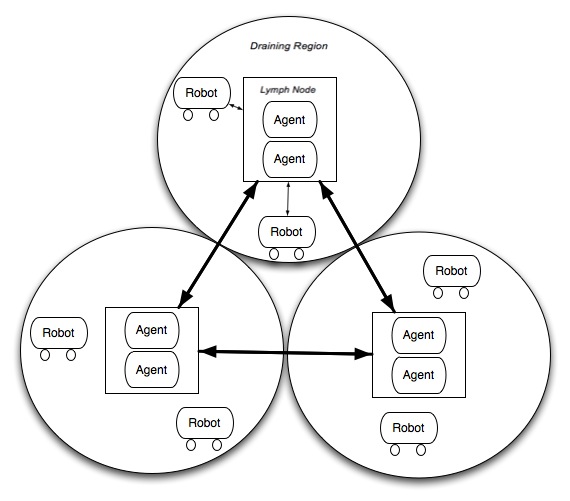}\\
\end{tabular}
\caption{A scaled up multi-robot AIS system with sub-modular architecture (Model 3). Bold arrows denote communication between servers and light arrows communication between robots and servers} 
\label{fig:goslim}
\end{figure*}

\begin{figure*}
\centering
\scriptsize
\begin{tabular}{l}
\includegraphics[scale=0.45]{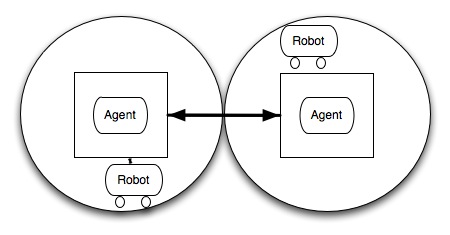}\\
\end{tabular}
\caption{A scaled down version of the multi-robot AIS system with sub-modular architecture (Model 3)} 
\label{fig:goslim}
\end{figure*}
	
	We are interested in an architecture that minimizes the time taken by a robot to transmit information about an obstacle, the time taken by an agent to transmit back an initial rule-set of actions and the time taken by an agent to communicate a new rule-set to all other agents. There are two potential communication bottlenecks (communication between robots and computer servers, and communication between computer servers) and four ways in which these resources can be constrained:
	
	\begin{enumerate}
	\item Unlimited Robot Bandwidth, Unlimited Server Bandwidth: Assuming that robots have unlimited bandwidth to communicate with computer servers and software agents can communicate with each other over a channel with unlimited bandwidth, we see that trivially any of the architectures would suffice.
	\item Limited Robot Bandwidth, Unlimited Server Bandwidth: Assuming communication between robots and computer servers is a bottleneck mandates a small fixed size DR i.e. a computer server servicing a small number of robots to reduce contention and transmission time. Since communication between servers is not constrained, we can have the number of servers scaling linearly with system size, giving Model 1 (completely modular network) as the optimal architecture.
	\item Unlimited Robot Bandwidth, Limited Server Bandwidth: Assuming communication between computer servers is a bottleneck stipulates a fixed number of computer servers to reduce communication overhead. Since communication between robots and servers is not constrained, we can have the DR size (number of robots serviced by a single server) and LN size (number of agents in a single server) scaling with system size. Hence the optimal architecture is Model 2 (non-modular detection network).
	\item Limited Robot Bandwidth, Limited Server Bandwidth: A bottleneck in robot and server communication demands a small DR and lots of servers, whereas a bottleneck in server communication requires a large server with fewer total number of servers. The architecture which balances these opposing requirements is Model 3 (hybrid sub-modular architecture) i.e. the number of servers and their size (number of robots serviced by each server) increases with system size.
	\end{enumerate}
	
	Understanding the tradeoff between fast search for pathogens and fast production of antibodies is important for AIS that mimic the NIS. If the goal of an AIS is only search or detection in physical space with a local response, then a completely modular design (Model 1) will be optimal. If an AIS searches in physical space but requires a global response after detection, a sub-modular architecture (Model 3) optimizes the tradeoff between local search and global response. Our analysis sheds light on the relationship between physical space and architecture in resource-constrained distributed systems.

\section{Concluding Remarks and Future Work}
The IS is comprised of rare antigen-specific IS cells that it must utilize to search for initially small numbers of pathogens localized in a large physical space. The IS solves this classic search for a needle in a haystack in time that is almost invariant of the size of the organism. The decentralized nature of the lymphatic network also helps in efficient pathogen detection by acting as a small volume of tissue where DCs can efficiently present antigen to T cells. The IS must also respond to the antigen by producing antibodies (in the case of WNV) proportional to the mass of the organism. From empirical data, that time also appears independent of body size.

We examined three different hypothesized IS architectures to explain the scale-invariant detection and response times of the IS. The sub-modular detection network strikes a balance between the two opposing goals of antigen detection (local communication) and antibody production (global communication), and is consistent with observed numbers and sizes of LN. 

The mechanisms used by the NIS to overcome physical constraints of system size are worthy of replication in AIS domains. Similar to NIS, physical space and resource are also important constraints on AIS, especially distributed systems applications used to connect low-powered sensors using short-range wireless communication. A sub-modular architecture is also optimal for AIS problems like distributed robot control that require a tradeoff between local search for a solution and global response or proliferation of the solution between different components.

\section{Acknowledgements}
  This work was supported by an NIH COBRE CETI grant (RR018754) to Dr. Melanie Moses and travel grants from PIBBS (Program in Interdisciplinary Biological and Biomedical Sciences) to Soumya Banerjee. The authors wish to thank Dr. Alan Perelson, Dr. Stephanie Forrest, Dr. Jed Crandall, Dr. Rob Miller, Dr. Jeremie Guedj, Kimberly Kanigel and Drew Levin for helpful discussions and Dr. Nicholas Komar for sharing his experimental data.  
  
\section{References}

[1]  Kleinberg, J. (2007), The Wireless Epidemic, Nature, Vol. 449, 287-288.

[2] Banerjee S, Moses M (2009) A Hybrid Agent Based and Differential Equation Model of Body Size Effects on Pathogen Replication and Immune System Response. In: P.S. Andrews et al. (Eds.) Artificial Immune Systems, 8th International Conference, ICARIS 2009, Lecture Notes in Computer Science, Springer Verlag, Berlin, Germany, vol 5666, pp 14-18

[3]  Banerjee, S. and Moses, M. (2010). Scale Invariance of Immune System Response
Rates and Times: Perspectives on Immune System Architecture and Implications for Artificial Immune Systems. Swarm Intelligence (in press)

[4] Banerjee S, Moses M (2010) Modular RADAR: An Immune System Inspired Search and Response Strategy for Distributed Systems. In: E. Hart et al. (Eds.) Artificial Immune Systems, 9th International Conference, ICARIS 2010, Lecture Notes in Computer Science, Springer Verlag, vol 6209, pp 116-129

[5]  Michael S. Diamond, et al. (2003). A Critical Role for Induced IgM in the Protection against West Nile
Virus Infection. Journal of Experimental Medicine 10.1084/jem20031223

[6]  Michel L. Bunning, et al. (2002). Experimental Infection of Horses With West Nile virus. Emerg. Infect.
Dis., 380(8):380-386

[7]  Soderberg, A.K. et al., (2005), Innate Control of Adaptive Immunity via Remodeling of Lymph Node
Feed Arteriole, PNAS, Vol. 102, 16315-16320.

[8]  Komar, N., S. Langevin, S. Hinten, N. Nemeth, E. Edwards, D. Hettler, B. Davis et al. 2003. Experimental infection of North American birds with the New York 1999 strain of West Nile virus. Emerg Infect Dis 9:311-322.

[9]  Nair, Shivashankar B. et al., (2008), An Immune System based Multi-Robot Mobile Agent Network,
The 7th International Conference on Artificial Immune Systems (ICARIS), Volume 5132-014, 424-433

[10]  Peters, R.H. (1983). The Ecological Implications of Body Size. Cambridge University Press, Cambridge

{
\footnotesize
\bibliographystyle{abbrv}
\addtolength{\itemsep}{-2mm}
\bibliography{mbsearch}
}
\end{document}